\documentclass[12pt,preprint]{aastex}
 \usepackage{emulateapj5}
 \usepackage{graphicx}
\usepackage{longtable}
\usepackage{ulem}
\usepackage{float}
\slugcomment{Accepted by ApJ, printed on \today}

\newcommand{\msun}{{M}_{\sun}}

\newcommand{\be}{\begin{equation}}
\newcommand{\ee}{\end{equation}}

\shorttitle{Jet Formation and Hot Plasma in AGNs}
\shortauthors{Wu et al.}

\font\mbf = cmmib10 scaled\magstep1
       \font\mbfs = cmmib10 \font\mbfss = cmmib10 scaled 833
\begin{document}

\title{A Physical Link Between Jet Formation and Hot Plasma in Active Galactic Nuclei}

\author{Qingwen Wu\altaffilmark{1}, Xinwu Cao\altaffilmark{2}, Luis C. Ho\altaffilmark{3} and Ding-Xiong Wang\altaffilmark{1}}

\altaffiltext{1}{School of Physics, Huazhong University of Science and Technology,
 Wuhan 430074, China; Email: qwwu@hust.edu.cn; dxwang@hust.edu.cn}

\altaffiltext{2}{Key Laboratory for Research in Galaxies and
Cosmology, Shanghai Astronomical Observatory, Chinese Academy of
Sciences, 80 Nandan Road, Shanghai, 200030, China; cxw@shao.ac.cn}

\altaffiltext{3}{The Observatories of the Carnegie Institution for Science,
813 Santa Barbara Street, Pasadena, CA 91101, USA; lho@obs.carnegiescience.edu }

\begin{abstract}
Recent observations suggest that in black hole X-ray binaries jet/outflow
formation is related to the hot plasma in the vicinity of the black hole,
either in the form of an advection-dominated accretion flow at low accretion
rates or in a disk corona at high accretion rates.  We test the viability of
this scenario for supermassive black holes using two samples of active galactic
nuclei distinguished by the presence (radio-strong) and absence (radio-weak) of
well-collimated, relativistic jets.  Each is centered on a narrow range of black
hole mass but spans a very broad range of Eddington ratios, effectively
simulating, in a statistical manner, the behavior of a single black hole
evolving across a wide spread in accretion states.  Unlike the relationship
between the radio and optical luminosity, which shows an abruptly break between
high- and low-luminosity sources at an Eddington ratio of $\sim 1\%$, the radio
emission---a measure of the jet power---varies continuously with the hard X-ray
(2--10 keV) luminosity, roughly as $L_{\rm R} \propto L_{\rm X}^{0.6-0.75}$.
This relation, which holds for both radio-weak and radio-strong active galaxies,
is similar to the one seen in X-ray binaries.
Jet/outflow formation appears to be closely linked to the conditions that give
rise to the hot, optically thin coronal emission associated with accretion
flows, both in the regime of low and high accretion rates.

\end{abstract}


\section{Introduction}
The ratio of radio to optical flux in active galactic nuclei (AGNs) spans a
very wide distribution, and is possibly bimodal; radio-loud (RL) AGNs are about
$10^{3}-10^{4}$ times brighter in the radio than radio-quiet (RQ)
AGNs with the same optical emission \citep[e.g.,][]{ke94,xu99}.
The physical origin of the RL/RQ
dichotomy is still an unresolved problem in AGN physics. One
possible explanation is that RL AGNs are driven by fast-rotating
black holes (BHs) through the Blandford-Znajek mechanism
\citep[][]{bz77}, while BHs in RQ AGNs are non-spinning or
spinning slowly. A radio-loudness parameter of $R\equiv F_{\rm 5\
GHz}/F_{B}\simeq 10$ is usually taken as the division between bright RL and
RQ quasars, where $F_{\rm 5\ GHz}$ is the monochromatic
flux density at 5~GHz and $F_{B}$ is the optical $B$-band flux
density at 4400~\AA\ \citep{ke94}. Extending the above $R$
parameterization to low-luminosity AGNs (LLAGNs), however, is not
straightforward, as most LLAGNs qualify as RL
according to the above $R$ criterion when their nuclear emission at
radio and optical wavelengths are considered \citep[][]{hp01,ho02}, even though
the majority of LLAGNs generally have very low radio powers and do not have
well-collimated, large-scale jets (Ho 2008).
Thus, the traditional definition of radio-loudness ($R=10$)
cannot distinguish LLAGNs from classical RL AGNs such as Fanaroff \& Riley
(1974, FR) type I and II sources.  \citet{si07} proposed that the critical
value of $R$ that divides RL and RQ sources may depend on Eddington ratio,
$L_{\rm bol}/L_{\rm Edd}$, where $L_{\rm bol}$ is the bolometric luminosity
and the Eddington luminosity $L_{\rm Edd}=1.28\times10^{38} (M_{\rm BH}/\msun)
\rm\ erg\ s^{-1}$.  To
avoid possible confusion regarding the definition of RL versus RQ, in this
work we simply use the terms ``radio-strong'' (RS) and ``radio-weak'' (RW)
to distinguish between sources that have highly relativistic, well-collimated
jets (e.g., blazars, FR I and FR II radio galaxies, and classical RL quasars)
and from those that do not (e.g., Seyfert galaxies and classical RQ quasars),
regardless of Eddington ratios or luminosities.

AGNs are powered by accretion of matter onto BHs, as are X-ray binaries (XRBs).
Both the optical/UV bumps observed in quasars and the soft X-ray emission
observed in the high/soft (HS) state of XRBs can be naturally interpreted as
multi-temperature blackbody emission from a cold, optically thick,
geometrically thin standard accretion disk \citep[SSD,][]{ss73}.  Models of
hot, optically thin, geometrically thick advection-dominated accretion flows
(ADAFs; also called radiatively inefficient accretion flows) have been
developed for BHs accreting at low mass accretion rates (e.g., Ichimaru 1977;
Narayan \& Yi 1994, 1995; Abramowicz et al. 1995; see Kato et al. 2008 and
Narayan \& McClintock 2008 for recent reviews). The ADAF model can successfully
explain most observational features of nearby LLAGNs and XRBs in their LH
state (e.g., Quataert et al. 1999; Yuan et al. 2005; Wu et al.
2007; Ho 2009; Yu et al. 2011; Xu 2011; Nemmen et al. 2012; Liu \& Wu 2013; see
reviews in Remillard \& McClintock 2006, Done et al. 2007, Yuan 2007, Ho 2008).
For Eddington ratios lower than a critical value of $\sim 1\%$, the hard X-ray
(2--10 keV) photon indices of both XRBs and AGNs are inversely correlated with
the Eddington ratio; the trend reverses for sources with Eddington ratios
$\gtrsim 1\%$ \citep[e.g.,][and references therein]{wg08,gc09,sp09,yp11}. The
different correlations between X-ray spectral slope and Eddington ratio
found in high-luminosity and low-luminosity sources are roughly consistent with
the expectations from the SSD/corona model \citep[][]{ca09} and the ADAF model
\citep[][]{ql13}, respectively. They provide strong evidence for accretion mode
transitions in BH accreting systems. Accretion mode transitions also offer
a plausible explanation for the trends between jet power and BH mass observed
in different classes of powerful RL AGNs \citep[e.g.,][]{gc01,wc08,xu09}.

Simultaneous multi-wavelength observations of BH XRBs in their LH state have
shown that their radio luminosity is correlated with their X-ray luminosity,
roughly as $L_{\rm R}\propto L_{\rm X}^{0.6}$ \citep[][]{co03,ga03}. Once the
sources enter the HS state, their radio emission becomes quenched, such that
$L_{\rm R}$ is significantly depressed at a given $L_{\rm X}$
\citep[e.g.,][]{fe99,co00}. The observed radio properties of XRBs suggest that,
apart from the possible influence of BH spin, jet formation is somehow also
regulated by accretion rate.  In an important recent study, \citet{zd11} found
that the radio$-$X-ray correlation in the BH XRB Cygnus X-1 extends to the HS
state only if its hard X-ray emission is considered (after subtracting the
blackbody component due to the SSD). Their result suggests that the radio
emission is related only to the hot plasma in accretion flows (e.g., ADAF or
corona above the SSD). In such a scenario, the suppression of radio emission
during in the HS state in XRBs may be caused by a reduction of the corona
resulting from its being cooled by the high luminosity of the SSD in the HS
state.  A similar phenomenon is seen in AGNs: the ratio coronal emission
(represented by the hard X-rays) to the bolometric luminosity decreases with
increasing Eddington ratio \citep[e.g.,][]{wa04,ca09}.  Thus, the hot
plasma---either in the form of an ADAF or a corona---may play a key role in
regulating the radio emission in both XRBs and AGNs.

\citet{zd11} proposed that the hot plasma (ADAF/corona), not the cold SSD,
controls the formation of a jet or outflow.  One practical difficulty in
testing this hypothesis further is that radio emission is too weak to be
detected in most XRBs in their HS state.  By contrast, radio data are much more
readily available for supermassive BHs, across a wide range of accretion rates.
Indeed, the existence of the so-called BH fundamental plane relation linking
radio and X-ray emission to BH mass suggests that the physics of accretion and
jet formation can be unified across a very wide mass range, stretching from
XRBs to AGNs \citep[][]{mhd03,fa04}.  In light of this, it should be possible to
use AGNs to test Zdziarski's hot plasma scenario for jet/outflow formation.
However, it should be noted that differences in BH mass or accretion rate alone
still cannot fully explain the radio dichotomy in AGNs; other parameters, such
as BH spin or the strength of the large-scale magnetic field, may also play a
key role \citep[e.g.,][]{sp10,bf11}.

We carefully assemble an AGN sample to test the viability of the hot plasma
jet/outflow formation.  A $\rm \Lambda CDM$ cosmology with $H_{0}=70\
\rm km\ s^{-1}\ Mpc^{-1}$, $\Omega_m=0.27$, and $\Omega_{\Lambda}=0.73$
is adopted in this work.

\begin{table*}[t]
   \centering
\begin{minipage}{144mm}

\footnotesize
  \centerline{\bf Table 1. Data for RW AGNs}
  \begin{tabular}{llcccccc}\hline\hline
Name & $z$ or $D_L$$^{\rm a}$ & log $L_{\rm 2-10keV}$ & log $L_{\rm 5\ GHz}$ & log $L_{B}$ & log $R$ & log $M_{\rm BH}$ & Refs.$^{\rm b}$  \\
         &    (Mpc)   &  ($\rm erg s^{-1}$) & ($\rm erg s^{-1}$) & ($\rm erg s^{-1}$) &  & ($\msun$) &   \\
\hline
Ark 120         &  0.0323  &  43.95 &  38.56 &  44.17 & -5.61  &  8.27  & 1,2,2,2  \\
Ark 374         &  0.0630  &  43.49 &  38.67 &  44.26 & -5.59  &  7.86  & 1,2,2,2  \\
Fairall 9       &  0.0470  &  43.97 &  39.11 &  44.52 & -5.41  &  7.91  & 1,2,2,2  \\
H0557-385       &  0.0339  &  44.03 &  39.42 &  43.51 & -4.09  &  7.64  & 1,2,2,2  \\
MC-5-23-16      &  0.0085  &  43.02 &  37.68 &  42.83 & -5.15  &  7.85  & 1,2,2,2  \\
Mrk 79          &  0.0222  &  43.12 &  38.35 &  43.67 & -5.32  &  8.01  & 1,2,2,2  \\
Mrk 279         &  0.0305  &  43.50 &  38.93 &  43.78 & -4.85  &  7.62  & 1,2,2,2  \\
Mrk 290         &  0.0296  &  43.25 &  38.34 &  43.85 & -5.51  &  7.65  & 1,2,2,2 \\
Mrk 509         &  0.0344  &  44.68 &  38.83 &  44.61 & -5.78  &  7.86  & 1,2,2,2  \\
Mrk 841         &  0.0364  &  43.89 &  38.18 &  44.20 & -6.02  &  7.88  & 1,2,2,2  \\
Mrk 1014        &  0.1630  &  43.85 &  40.45 &  44.86 & -4.41  &  8.03  & 1,2,2,2  \\
NGC 2992        &  0.0077  &  42.97 &  38.83 &  43.67 & -4.84  &  7.72  & 1,2,2,2  \\
NGC 5548        &  0.0172  &  43.39 &  38.70 &  43.60 & -4.90  &  8.03  & 1,2,2,2  \\
NGC 7213        &  0.0060  &  42.27 &  38.96 &  43.18 & -4.22  &  7.99  & 1,2,2,2  \\
PG 0052+251     &  0.1550  &  44.61 &  39.50 &  45.01 & -5.51  &  8.40  & 1,2,2,2  \\
PG 0804+761     &  0.1000  &  44.46 &  39.40 &  44.89 & -5.49  &  8.24  & 1,2,2,2  \\
PG 0953+414     &  0.2341  &  44.73 &  40.17 &  45.57 & -5.40  &  8.24  & 1,2,2,2  \\
PG 1048+342     &  0.1670  &  44.04 &  37.57 &  44.87 & -7.30  &  8.37  & 1,2,2,2  \\
PG 1115+407     &  0.1550  &  43.93 &  38.98 &  44.73 & -5.75  &  7.67  & 1,2,2,2  \\
PG 1307+085     &  0.1550  &  44.08 &  39.10 &  45.17 & -6.07  &  7.90  & 1,2,2,2  \\
PG 1322+659     &  0.1680  &  44.02 &  38.88 &  44.97 & -6.09  &  8.28  & 1,2,2,2  \\
PG 1402+261     &  0.1640  &  44.15 &  39.56 &  45.02 & -5.46  &  7.94  & 1,2,2,2  \\
PG 1415+451     &  0.1140  &  43.60 &  38.82 &  44.18 & -5.36  &  8.01  & 1,2,2,2  \\
PG 1427+480     &  0.2210  &  44.20 &  38.14 &  44.51 & -6.37  &  8.09  & 1,2,2,2  \\
PG 1501+106     &  0.0364  &  43.69 &  39.00 &  44.20 & -5.20  &  7.88  & 1,2,2,2  \\

\hline
NGC  224 (M31)  &   0.7    &  35.85 &  32.10 &   ---  & ---   &   7.82  & 3,4,5  \\
NGC  266        &  62.4    &  40.87 &  37.78 &   ---  & ---   &   8.37  & 6,7,5  \\
NGC  821        &  23.2    &  38.30 &  35.40 &   ---  & ---   &   8.21  & 8,8,5  \\
NGC 1097        &  14.5    &  40.63 &  36.29 &   ---  & ---   &   8.08  & 6,7,5  \\
NGC 1667        &  61.2    &  40.55 &  37.40 &  40.86 & -3.46 &   7.81  & 3,9,10,5  \\
NGC 2787        &  13.0    &  38.31 &  36.40 &  38.90 & -2.50 &   8.14  & 11,12,4,5  \\
NGC 2841        &  12.0    &  38.30 &  36.00 &  39.34 & -3.34 &   8.31  & 8,8,10,5  \\
NGC 3031        &  1.4     &  39.41 &  36.21 &  39.41 & -3.20 &   7.73  & 11,12,4,5  \\
NGC 3147        &  40.9    &  41.87 &  38.01 &  40.52 & -2.51 &   8.29  & 3,9,10,5 \\
NGC 3169        &  19.7    &  41.05 &  37.20 &  39.39 & -2.19 &   8.01  & 3,4,10,5,  \\
NGC 3245        &  22.2    &  39.29 &  36.98 &  40.06 & -3.08 &   8.21  & 3,4,4,5  \\
NGC 3379        &  8.1     &  37.53 &  35.73 &  38.70 & -2.97 &   8.18  & 3,4,4,5 \\
NGC 4143        &  17.0    &  40.03 &  37.18 &  41.15 & -3.97 &   8.16  & 3,7,10,5  \\
NGC 4168        &  16.8    &  38.97 &  36.93 &  38.96 & -2.03 &   7.97  & 13,9,14,5  \\
NGC 4203        &  9.7     &  40.80 &  36.70 &  39.80 & -3.10 &   7.79  & 15,12,4,5  \\
NGC 4216        &  16.8    &  38.91 &  36.79 &  ---   & ---   &   8.09  & 3,16,5\\
NGC 4235        &  35.1    &  42.25 &  37.57 & 40.87  & -3.30 &   7.60  & 13,9,10,5 \\
NGC 4459        &  16.8    &  38.87 &  36.09 &  39.30 & -3.21 &   7.82  & 3,4,4,5  \\
NGC 4477        &  16.8    &  39.60 &  35.64 &  39.58 & -3.94 &   7.89  & 3,9,10,5 \\
NGC 4501        &  16.8    &  39.59 &  36.25 &  39.62 & -3.37 &   7.79  & 3,9,10,5 \\
NGC 4579        &  16.8    &  41.51 &  37.81 &  40.72 & -2.91 &   7.77  & 3,9,9,5 \\
NGC 4621        &  16.8    &  37.80 &  35.10 &   ---  & ---   &   8.34  & 17,8,5  \\
NGC 4636        &  17.0    &  39.38 &  36.4  &  39.90 & -3.50 &   8.14  & 3,16,10,5  \\
NGC 4697        &  12.4    &  37.30 &  35.00 &   ---  & ---   &   7.83  & 17,8,5  \\
NGC 5033        &  18.7    &  41.08 &  38.00 &  41.40 & -3.40 &   7.60  & 13,9,10,5 \\
NGC 7582        &  22.3    &  41.69 &  38.55 &   ---  & ---   &   7.67  & 18,8,5  \\
NGC 7603        &  124.2   &  43.65 &  39.00 &  42.10 & -3.10 &    8.14 & 19,4,4,5  \\
\hline
\end{tabular}
\tablecomments{
$^{\rm a}$ The distances of nearby LLAGNs are adopted from \citet{ho09}.\\
$^{\rm b}$ References for X-ray, radio, and optical luminosities, and BH mass, respectively:
(1) \citet{zz10}; (2) \citet{bi09} ; (3) \citet{ho09}; (4) \citet{ho02}; (5) the BH mass is estimated from the
$M_{\rm BH}-\sigma_{*}$ relation of G\"ultekin et al. (2009), adopting $\sigma_{*}$ values from \citet{ho09}
except for four sources (NGC 1097, NGC 4697, NGC 7582, and NGC 7603), which were taken from the Hyperleda database;
(6) \citet{er10}; (7) \citet{ne13}; (8) \citet{yu09}; (9) \citet{hu01}; (10) the optical $B$-band luminosity is estimated
from the total H$\beta$ luminosity in this work; (11)\citet{ho01}; (12) \citet{si07}; (13) \citet{pa06};
(14) \citet{hp01}; (15) \citet{pi10}; (16) \citet{na05}; (17) \citet{wr08}; (18) \citet{gk09}; (19) \citet{ma07}.
}

\end{minipage}
\end{table*}

\section{Sample}

To allow for the possibility of a radio dichotomy, we explore the relation
between jets and accretion processes separately for RS and RW sources.  We
impose two selection criteria for each sample: (1) that the sources span a
narrow range of BH masses ($\pm$0.4 dex) and (2) that they cover as wide a
range of luminosities as possible to mimic a large dynamic range in accretion
rate or Eddington ratio at a fixed mass.  The spread of 0.4 dex in BH mass was
chosen to reflect the typical uncertainty in current methods of estimating
$M_{\rm BH}$, either from broad emission lines for type 1 AGNs \citep[e.g.,][]{vp06}
or from the $M_{\rm BH}-\sigma_*$ \citep[e.g.,][]{gk09,ho09} or
$M_{\rm BH}-L_{\rm bulge}$ \citep[e.g.,][]{mh03} relation for type 2 AGNs and
very low-luminosity AGNs. We note that our samples are selected from the
literature and are quite heterogeneous and incomplete.  Nevertheless, the main
conclusions of this work do not depend on the statistical completeness of the
sample.  We describe the RW and RS samples separately below.

\subsection{RW AGNs}
The RW sample, summarized in Table 1, includes both bright, low-redshift type 1
AGNs (classical Seyfert 1s and quasars) and nearby LLAGNs (low-luminosity
Seyferts and LINERs) without evident large-scale, well-collimated jets. We
construct a sample of 52 RW AGNs with BH masses lying in the narrow range
$M_{\rm BH} = 10^{8\pm0.4}\,\msun$, using the AGN catalogs from CAIXA
\citep[][]{bi09} and the Palomar spectroscopic survey of nearby galaxies
\citep[][]{ho95}. These samples have well-determined multiwavelength
observations and BH mass estimates. The average BH mass, $10^{8}\,\msun$,
is chosen to optimize both the size and the luminosity coverage of the sample.

We select 25 optically bright type 1 AGNs from the CAIXA catalog, for which
BH masses, radio core flux densities, and absolute $V$ magnitudes are available
\citep[][and references therein]{bi09}.  The BH masses of these type 1 AGNs are
calculated from the virial product of the velocity widths of the broad H$\beta$
emission line and the size of the broad-line region estimated from the
size-luminosity relation of reverberation-mapped AGNs \citep*[][]{kas00,kas05};
typical uncertainties in the mass estimates are $\sim 0.4$ dex
\citep*[e.g.,][]{vp06,mcgill08}.  We obtain $B$-band luminosities from the
$V$ absolute magnitudes and $B-V$ colors given in \citet{bi09};
we assume $B-V=0.3$ mag if no color information is available.  We additionally
make use of 2--10 keV X-ray luminosities from \citet{zz10}, which are derived
from targeted observations from {\it XMM-Newton}.

We add an additional 27 LLAGNs from the literature to increase the luminosity
coverage of the sample. Most of the sources are taken from the Palomar survey
\citep[][]{ho95}, which has well-determined multiwavelength data, including
interferometric radio observations \citep*[e.g.,][]{hu01,ho02,na05} and X-ray
observations from {\it Chandra}\ or {\it XMM-Newton}\
\citep*[e.g.,][]{ho01,pa06,ho09}.  As emphasized in Ho (2008), high-resolution
data are essential to properly isolate the low-level nuclear component of the
LLAGN from the often much brighter emission from the surrounding host galaxy.
Since high-resolution optical continuum measurements are not uniformly
available for the Palomar galaxies, we estimate the optical luminosities of the
LLAGNs from their H$\beta$ line luminosity, using the correlation between
$L_{\rm H\beta}$ and $M_{B}$ for both bright and faint type 1 AGNs, as
calibrated by \citet{hp01}.  We take the H$\beta$ luminosities directly
from \citet{ho02} when available; otherwise, we derive H$\beta$ from the
extinction-corrected total H$\alpha$ luminosity (narrow plus broad component,
if present) and the ratio of H$\alpha$ to H$\beta$ \citep{ho97a,ho97b,ho03}.
We estimate BH masses for the LLAGNs from the stellar velocity dispersion
$\sigma_{*}$ of the host galaxy bulge, using the $M_{\rm BH}-\sigma_{*}$
relation of \citet{gk09}, which has an intrinsic scatter of $\sim 0.4$ dex.
Velocity dispersions of 23 sources are available from \citet{hg09}, and the
remaining four were found in the Hyperleda
database\footnote{{\tt http://leda.univ-lyon1.fr}} (see Table 1 for details).

\begin{table*}[t]
\footnotesize
   \centering
 \begin{minipage}{144mm}
  \centerline{\bf Table 2. Data for RS AGNs (FR Is and FR IIs)}
  \begin{tabular}{lcccccc}\hline\hline
Name & $z$  &Type & $\log L_{\rm 2-10keV}$ & $\log L_{\rm 178\ MHz}$
& $\log M_{\rm BH}$ & Refs. \\
         &   &       & ($\rm erg s^{-1}$) &  ($\rm erg s^{-1}$) & ($\msun$) &  \\
\hline
3C 31    &  0.0167  &  I  &   40.67   & 40.31   & 8.70  &  1   \\
3C 48    &  0.3670  &  I  &   45.00   & 43.64   & 9.20  &  1    \\
3C 66B   &  0.0215  &  I  &   41.10   & 40.69   & 8.84  &  1   \\
3C 83.1B &  0.0255  &  I  &   41.13   & 40.88   & 9.01  &  1   \\
3C 84    &  0.0177  &  I  &   42.91   & 40.92   & 8.64  &  1   \\
3C 264   &  0.0208  &  I  &   41.87   & 40.69   & 8.80  &  1   \\
3C 272.1 &  0.0029  &  I  &   39.35   & 38.84   & 8.80  &  1   \\
3C 296   &  0.0237  &  I  &   41.49   & 40.51   & 8.80  &  1   \\
3C 338   &  0.0303  &  I  &   42.31   & 41.29   & 8.92  &  1   \\
3C 346   &  0.1620  &  I  &   43.40   & 42.15   & 8.89  &  1   \\
3C 442A  &  0.0270  &  I  &   41.10   & 40.71   & 8.40  &  1    \\
3C 449   &  0.0171  &  I  &   40.35   & 40.16   & 8.54  &  1   \\
3C 465   &  0.0293  &  I  &   41.04   & 41.16   & 9.13  &  1    \\
NGC 6251 &  0.0240  &  I  &   41.60   & 40.43   & 8.98  &  1   \\
\hline
3C 20    &  0.1740  &  II &   44.05   & 42.82   & 8.66  &  2   \\
3C 33.1  &  0.1810  &  II &   44.38   & 42.34   & 8.58  &  2   \\
3C 47    &  0.4250  &  II  &  45.05   & 43.52   & 9.20  &  2   \\
3C 79    &  0.2559  &  II &   44.18   & 43.07   & 8.96  &  2   \\
3C 171   &  0.2384  &  II &   44.08   & 42.80   & 8.70  &  2   \\
3C 184.1 &  0.1187  &  II &   43.91   & 41.95   & 8.57  &  2   \\
3C 284   &  0.2394  &  II &   43.98   & 42.57   & 9.09  &  2   \\
3C 285   &  0.0794  &  II &   43.33   & 41.53   & 8.61  &  2   \\
3C 349   &  0.2050  &  II &   43.87   & 42.48   & 8.75  &  2   \\
3C 436   &  0.2145  &  II &   43.53   & 42.65   & 9.06  &  2   \\
3C 452   &  0.0811  &  II &   44.00   & 42.23   & 8.69  &  2   \\
3C 457   &  0.4280  &  II &   44.56   & 43.23   & 9.09  &  3   \\
\hline
\end{tabular}
\tablecomments{
References for BH mass: (1) \citet{wu11}; (2) estimated in this work; (3) \citet{mc04}.
}
\end{minipage}

\end{table*}

\subsection{RS AGNs}
In order to explore the relation between relativistic jets and accretion
processes in RS AGNs, it is important to isolate emission radiated separately
from the jet and the disk, since the disk emission can be contaminated by the
powerful jets.  The nuclear emission in RS AGNs can also be Doppler boosted for
relativistic jets viewed at small angles (e.g., blazars); conversely, it can be
even deboosted for relativistic jets with large viewing angles (e.g., FR I/IIs).
For the purpose of this work, we choose FR I and FR II radio galaxies, which
are viewed at relatively large angles. In these sources the contribution of
jet emission at X-ray energies should not be important due to the deboosting
effect. For example, for a jet viewing angle of $45^{\rm o}$ to $90^{\rm o}$,
the jet emission can be deboosted by several tens to a thousand times for a
typical jet Lorentz factor of $\sim 10$ \citep[e.g.,][]{ge93}. Recent studies
also suggest that the X-ray emission of FR I/IIs are dominated by accretion
processes \citep[ADAFs in FR Is and disk corona in FR IIs;][]{gl03,wu07,ha09}.
To estimate the jet power, we use low-frequency radio emission because it
usually originates in diffuse, optically thin radio lobes moving at low
velocities, for which Doppler boosting or deboosting effects are negligible
\citep[e.g.,][]{wi99}.

We initially select FR I/II galaxies from the 3CRR catalog, most of which have
multi-waveband observations (e.g., low-frequency radio powers, host galaxy
luminosity, and narrow emission-line luminosity). \citet{ha09} collected data for
$\sim$100 sources in the 3CRR catalog to investigate the nature of the nuclei of
FR I/IIs that have been observed by {\it Chandra}\ or {\it XMM-Newton}.  The
X-ray spectra of these sources were fit by a two-component model if a single
power law is unable to provide a satisfactory fit \citep[for details,
see][]{ha09}. Following \citet{ha09}, we regard the heavily absorbed component
as the ``accretion-related'' component, and we adopt the absorption-corrected
X-ray luminosities in this work. We only choose FR I/IIs with well-measured
absorbing column density, leaving out those with uncertain $N_{\rm H}$ or
$N_{\rm H}>10^{24}$$\rm cm^{-2}$ to ensure that their intrinsic X-ray emission
in the 2--10 keV can be accurately measured. \citet{ba96} showed that
low-frequency radio emission may not be a good tracer of jet power in sources
with bright hot spots, as the emission of the lobes is dominated by the strong
radiation of the spots. We thus exclude FR IIs with previously reported
detections of hot spots in the X-ray or optical bands, as listed in
\citet[][and references therein]{ha04}.  The BH masses of the FR Is are
estimated from the stellar velocity dispersion of the host and the
$M_{\rm BH}-\sigma_{*}$ relation.  The FR IIs generally do not have measurements
of $\sigma_{*}$ available, and for these we obtain their BH masses from their
$H$-band luminosity \citep{bu10}, $M_{\rm BH} - L_H$ relation of \citet{mh03}.
The uncertainty of BH masses estimated from the bulge luminosity
is comparable to that estimated from the $M_{\rm BH}-\sigma_{*}$
relation \citep[$\sim 0.4\rm\ dex$; e.g.,][]{gk09}.

The final RS sample consists of 26 radio galaxies (14 FR Is and 12 FR IIs) with
$M_{\rm BH} = 10^{8.8\pm0.4}\, \msun$.  Again, the narrow range of mass was
chosen by design, to mimic, as closely as possible, a sample with fixed
$M_{\rm BH}$. Table 2 lists their radio
luminosities at 178~MHz, hard (2--10 keV) X-ray luminosities, and BH masses.
It should be noted that the RW and RS samples do not overlap in BH mass; the
latter has $M_{\rm BH}$ higher by 0.8 dex.  This is a consequence of the fact
that most FR I/IIs are hosted by giant elliptical galaxies, whereas RW AGNs
cover a broad range of galaxy types, both spirals and ellipticals, and hence
a broader range of $M_{\rm BH}$.  Unfortunately, the available data do not
permit us to construct a sample of RW and RS AGNs at fixed $M_{\rm BH}$ that
simultaneously covers a wide enough luminosity range for our purposes.  In what
follows, we will examine the possible connection between the hot plasma and
jets/outflows separately for the RW and RS samples.

\section{Results}
The sources in the RW sample have a fixed BH mass, $M_{\rm BH} = 10^{8.0\pm0.4}
\, \msun$, and cover a wide range of Eddington ratios,
$-8\lesssim\log L_{\rm bol}/L_{\rm Edd} \lesssim 0$,
assuming $L_{\rm bol} = 10 L_{B}$ (McLure \& Dunlop 2004).  By design, this
sample is meant to mimic a single AGN experiencing a wide range of accretion
rates, from very high to very low.
 \begin{figure}[H]
\begin{minipage}{90mm}
\includegraphics[width=90mm]{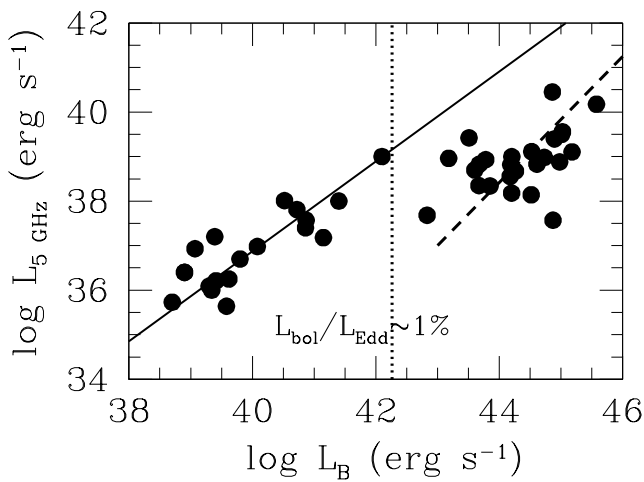}
\end{minipage}
\caption{Relation between $L_{B}$ and $L_{\rm 5\ GHz}$ for RW AGNs with BH masses in the range
$M_{\rm BH}=10^{8\pm0.4}\,M_\odot$. The dotted line corresponds to $L_{\rm
bol}/L_{\rm Edd}=1\%$, assuming $L_{\rm bol}=10 L_{B}$; the solid
 and dashed lines are the best fits for sources with $L_{\rm bol}/L_{\rm
Edd}$ below and above this limit, respectively.}
\label{f1}
\end{figure}
Figure 1 shows the relation between $L_{\rm 5\ GHz}$ and $L_{B}$.  We find
that optically luminous sources follow a different trend from weaker sources,
with the two populations roughly divided at a critical Eddington ratio of
$L_{\rm bol}/L_{\rm Edd} \approx 1\%$. The dashed and solid lines represent
the best-fitting linear relation between $L_{\rm 5\ GHz}$ and $L_{B}$ for
sources with Eddington ratios above and below this critical limit.  The
regression fits assume typical uncertainties of 0.2 and 0.3 dex for the radio
and optical luminosities, respectively (Ho \& Peng 2001).  For
$L_{5\ \rm GHz}\propto L^{\alpha}_{B}$, $\alpha=1.42\pm0.22$ for luminous
sources and $\alpha=1.01\pm0.10$ for weaker sources.  The two slopes are
significantly
different.  More importantly, the luminous sources with Eddington ratios above
the critical limit lie systematically offset below the best-fit line of the
fainter sources below this limit.  This can also be seen from examination of
the distribution of the radio-loudness parameter: the $R$ value of LLAGNs is
on average $\sim$2 orders of magnitude higher than that of luminous AGNs.
According to the Kolmogorov-Smirnov test, the hypothesis that the two
distributions of $R$ are drawn from the same parent population can be rejected
with a probability of $p<0.001$.

By contrast, for all but one source (M31) the 5 GHz radio luminiosity is
tightly correlated with the hard X-ray (2--10 keV) luminosity over 8 orders of
magnitude (Figure 2).  Excluding M31, the best-fit linear regression for the RW
sample yields

    \be
     \log L_{5\rm\ GHz}=(0.58\pm0.02)\log \it L_{\rm 2-10\ keV}+\rm(13.68\pm0.71),
     \ee
where we assume typical uncertainties of 0.2 and 0.3 dex for the radio and
X-ray luminosities, respectively \citep[e.g.,][]{hp01,st05}. To ensure that
the correlation between radio and X-ray luminosity is not spurious as a
consequence of the mutual dependence on distance, we performed a Spearman
partial correlation analysis using distance as the third variable; the
probability for accepting the null hypothesis that $L_{5\rm\ GHz}$ and
$L_{\rm 2-10\ keV}$ are uncorrelated is $p = 9.8\times10^{-6}$. We also find
that the radio and X-ray fluxes remain significantly correlated for RW AGNs,
with a Spearman correlation coefficient of $\rho=0.53$ and
$p=2.4\times10^{-5}$.  Both of these tests suggest that the
$L_{5\rm\ GHz}-L_{\rm 2-10\ keV}$ correlation is not driven by the distance
effect.

 \begin{figure}[H]
\begin{minipage}{90mm}
\includegraphics[width=90mm]{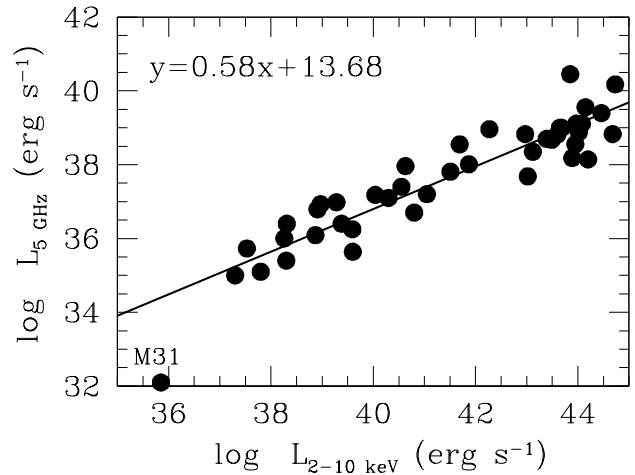}
\end{minipage}
\caption{Relation between $L_{\rm2-10\ keV}$ and $L_{\rm 5\ GHz}$ for RW AGNs with BH 
masses in the range $M_{\rm BH}=10^{8\pm0.4}\,M_\odot$.
 The solid line is the best fit for all sources, except for M31, which evidently
 deviates from other sources.}
\label{f2}
\end{figure}

The RS sample, anchored at $M_{\rm BH} = 10^{8.8\pm0.4}\, \msun$, also has a
wide distribution of Eddington ratios ($-6\lesssim \log L_{\rm bol}/L_{\rm Edd}
\lesssim -1$). As expected, the X-ray luminosities of FR IIs are
systematically higher than those of FR Is. Figure 3 shows the relation between
178 MHz and 2--10 keV luminosities; these two quantities are tightly
correlated over $\sim 5$ orders of magnitude in luminosity. The best fit,
once again assuming uncertainties of 0.2 and 0.3 dex for the
radio and X-ray luminosities, is

  \be
  \log L_{178\ \rm MHz}=(0.76\pm0.04)\log \it L_{\rm 2-10\ keV}+\rm (9.94\pm1.85).
  \ee
A Spearman partial correlation analysis confirms that this correlation is not
an artifact of a mutual dependence on distance ($p=4.3\times10^{-3}$),
although the radio and X-ray fluxes themselves are only correlated at a
mildly significant level ($\rho=0.38$ and $p=0.04$).

 \begin{figure}[H]
\begin{minipage}{90mm}
\includegraphics[width=90mm]{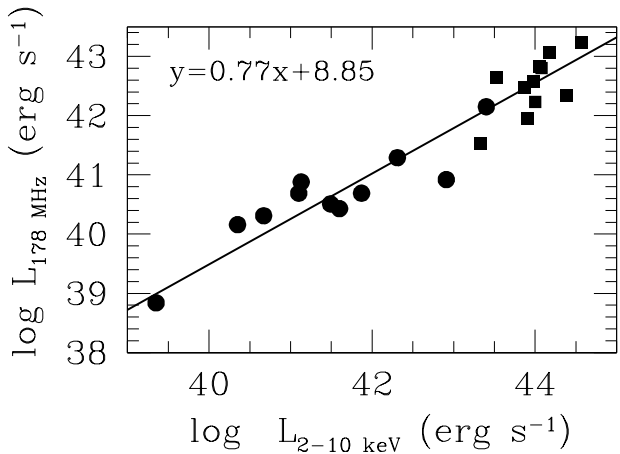}
\end{minipage}
\caption{Relation between $L_{\rm2-10\ keV}$
and $L_{\rm 178\ MHz}$ for FR I (circles) and FR II (squares) sources with BH
masses in the range $M_{\rm BH}=10^{8.8\pm0.4}\,M_\odot$.
The solid line is the best fit for all the sources.}
\label{f3}
\end{figure}

\section{Discussion}

Both highly collimated relativistic jets and less collimated nonrelativistic
outflows are common features in a variety of astrophysical objects, most
notably in BH XRBs and AGNs. Multi-waveband observations of BH XRBs suggest that
jet/outflow formation is closely correlated to the accretion mode in BH XRBs:
jets preferentially form in the ADAF state but are suppressed when the ADAF
transits to the SSD \citep[e.g.,][]{ga03,co03}.  In an interesting recent study,
\citet{zd11} found that the jet of the XRB Cygnus X-1 may be regulated by its
hot plasma. In this work, we investigate this issue using two samples of AGNs
with (RS) and without (RW) powerful relativistic jets.  Each of the two samples
is restricted to an essentially identical BH mass but spans a wide dynamic range
in Eddington ratio.  Our goal is to simulate, using a large sample of AGNs, as
closely as possible the condition of a {\it single}\ supermassive BH evolving
through a large variation in accretion rate.   How does accretion mode
transition affect the jet properties as traced through radio emission?

The optical continuum of bright RW AGNs originates mainly from the
multi-temperature blackbody emission from the SSD, whereas in LLAGNs it may
come from an ADAF.  Thus, the relation between the radio and optical luminosity
provides clues to the connection between the jet/outflow and the disk, a subject
investigated by many groups \citep[e.g.,][]{xu99,hp01,wlh04,si07}.  In RW AGNs,
the radio$-$optical correlation of high-luminosity sources clearly deviates
from that of LLAGNs (Figure 1); they separate near a dividing line of
$L_{\rm bol}/L_{\rm Edd} \simeq 1\%$, which roughly corresponds to the
theoretical prediction for the accretion mode to transition from an ADAF at low
luminosities to an SSD at high luminosities. Because the radiative efficiency
increases significantly when an ADAF switches to an SSD, the data points
for the bright AGNs shift toward higher optical luminosity than the LLAGNs.
Although Figure 1 gives the impression that the jet emission is strongly
suppressed in bright AGNs, we suggest that two effects, in fact, contribute to
this apparent offset: (1) an enhancement of the radiative efficiency as the
source evolves from the LH ADAF state to the HS SSD state; (2) a genuine
decrease in the efficiency of jet production at high luminosities (accretion
rates).  Both occur above a critical Eddington ratio threshold of
$L_{\rm bol}/L_{\rm Edd} \simeq 1\%$, and both manifest themselves as a
reduction of the radio-loudness parameter $R$.  But instead of viewing the
decline in $R$ as caused solely by a decrease in radio power, we suggest that
at least part of the relative decrease of radio to optical luminosity is
actually due to the increase of the optical disk
(optical/UV) emission in the HS state. Conversely, it is the reduction of the
disk emission during the LH ADAF state that is partly responsible for the
boost in $R$, an effect found in LLAGNs (Ho 1999; Ho \& Peng 2001; Ho
2002).  This can be seen in Figure 1 by the points that cluster around a
roughly constant radio luminosity of $L_{\rm 5\,GHz} \simeq 10^{38.5\pm1}\,
{\rm erg\,s^{-1}}$ and yet whose optical luminosities are spread over $\sim 4$
orders of magnitude ($L_B \approx 10^{41} - 10^{45}\,{\rm erg\,s^{-1}}$).

But this is not the whole story.  As in XRBs, we believe that the radio emission
in AGNs is additionally suppressed in the HS state.  Similar to the case of
the well-studied XRB Cygnus X-1 \citep{zd11}, AGNs, both RW (Figure 2) and
RS (Figure 3), obey a {\it single}, tight correlation between radio and hard
X-ray luminosity across an enormous range in Eddington ratios, which formally
includes systems in the LH and HS states.  Parameterizing the relation as
$L_{\rm R}\propto \it L_{\rm X}^{\alpha}$, $\alpha \approx 0.6$ for RW sources
and $\alpha \approx 0.75$ for RS sources.  In the case of Cygnus X-1,
Zdziarski et al. (2011; see also Miller et al. 2012) proposed that the
existence of a single radio$-$X-ray correlation that encompasses the LH and HS
states implies that the jet/outflow may be controlled primarily by the
hot plasma (ADAF or disk corona).  The analysis presented in this paper
extends this concept to supermassive BHs.

The hot plasma in the vicinity of the BH may couple naturally to the production
of jets and outflows for several reasons.  First, the optically thin, hot plasma
in either the ADAF or the corona reaches temperatures that are nearly virial
and has a positive Bernoulli parameter \citep[e.g.,][]{ny94,ny95}. Under these
conditions the material can escape as an outflow \citep[e.g.,][]{bb99,ca02,mf02}.
Second, jets and outflows are driven by the large-scale poloidal magnetic field
\citep[e.g.,][]{bz77,bp82}, which, according to the dynamo theory, has a length
scale comparable to the disk thickness \citep[e.g.,][]{me01}.  A hot ADAF or
corona, being quasi-spherical, has the geometric configuration conducive to
sustaining a strong poloidal magnetic field if it is in pressure equilibrium
with the thermal plasma.  Finally, the radio emission may originate from the
very same medium that produces the X-ray emission.  For example, the radio
emission may be synchrotron radiation from nonthermal electrons that are
generated by turbulence, magnetic reconnection, or weak shocks in the hot plasma
\citep[e.g.,][]{yu03,lb08,di10,liu13}. However, it is unclear whether this
coronal model can explain the level of radio emission observed in powerful
quasars; a forthcoming paper will address this issue.

Lastly, we note that the correlation between radio and hard X-ray emission
extends to RS AGNs, which in our study are represented by FR I and FR II radio
galaxies, the equivalent of ADAF-dominated and SSD-dominated sources,
respectively.  By analogy with the situation for the RW sources, even the
highly relativistic, well-collimated jets of FR I/II radio galaxies appear to
be connected to the hot plasma.  We do not have novel ideas to contribute to
the still unresolved physical origin for the existence of the radio dichotomy.
Parameters that may be important include the large-scale magnetic field (Cao
2011) and the BH spin (e.g., Wu et al. 2011; Tchekhovskoy et al. 2011; Li \&
Cao 2012; Narayan \& McClintock 2012, and see Spruit 2010 for a recent review).

\acknowledgments We thanks Feng Yuan, Bifang Liu and members of HUST astrophysics
group for many useful discussions and comments. This work was supported by the National Basic
Research Program of China (2009CB824800), the NSFC (grants
11143001, 11103003, 11133005, 11173043, 11121062, 10833002 and 11173011), the
Doctoral Program of Higher Education (20110142120037), and the CAS/SAFEA
International Partnership Program for Creative Research Teams (KJCX2-YW-T23).
LCH acknowledges support from the Carnegie Institution for Science, and
he thanks the Chinese Academy of Sciences and the hospitality of the National
Astronomical Observatories, where part of this work was done.

\end{document}